\documentclass[
    preprint,
    amsmath,
    amssymb,
    aps,
    physrev,
]{revtex4-2}

\usepackage{graphicx}
\usepackage{dcolumn}

\begin{document}
\preprint{preprint}

\title[Short title]{Precision Spectroscopy of the Fine Structure in the $\boldsymbol{a\;{}^3\Sigma_u^+(v=0)}$ and $\boldsymbol{c\;{}^3\Sigma_g^+(v=4)}$ States of the Helium Dimer}

\author{V. Wirth}
\affiliation{Institute of Molecular Physical Science, ETH Zurich, Zurich, Switzerland}%

\author{M. Holdener}%
\affiliation{Institute of Molecular Physical Science, ETH Zurich, Zurich, Switzerland}%

\author{F. Merkt}
\email{frederic.merkt@phys.chem.ethz.ch}
\affiliation{Institute of Molecular Physical Science, ETH Zurich, Zurich, Switzerland}%
\affiliation{Quantum Center, ETH Zurich, Zurich, Switzerland}
\affiliation{Department of Physics, ETH Zurich, Zurich, Switzerland}

\date{\today}

\begin{abstract}
With four electrons, He$_2$ is one of very few molecules for which first-principles quantum-chemical calculations that include the treatment of nonadiabatic, relativistic and quantum-electrodynamics corrections are possible. Precise spectroscopic measurements are needed as references to test these calculations. We report here on a spectroscopic measurement of the $c\;{}^3\Sigma_g^+\leftarrow a\;{}^3\Sigma_u^+$ electronic transition of $^4$He$_2$ at a precision ($\Delta \nu/\nu$) of $2.5\times 10^{-10}$ and with full resolution of the rotational, spin-spin and spin-rotational fine structures. 
The focus of the investigation is placed on the measurement and analysis of transitions to the rotational levels of the $c\;{}^3\Sigma_g^+(v=4)$ state, which are located energetically above the $\mathrm{He(1\;{}^1S_0) + He(2\;{}^3S_1)}$ dissociation limit and can decay by tunneling predissociation through a barrier in the potential of the $c$ state. 
The new data include a full set of the energy levels of the $a\;{}^3\Sigma_u^+(v=0)$ and $c\;{}^3\Sigma_g^+(v=4)$ states with rotational quantum numbers $N$ up to 9 and 10, respectively, and full sets of fine-structure intervals in these levels for comparison with the results of first-principles calculations carried out in a parallel investigation [B. R{\'a}csai, P. Jeszenszki, A. Marg{\'o}csy and E. M{\'a}tyus, arXiv:2506.23879v1 [physics.chem-ph] (2025)].
The new data were combined with data from earlier measurements of the spectrum of the $c-a$ band system of $^4$He$_2$ to derive full sets of molecular constants for the $a\;{}^3\Sigma_u^+(v=0)$ and $c\;{}^3\Sigma_g^+(v=4)$ states with much improved precision over previous experimental results. A pronounced broadening of the linewidths of the transitions to the $c\;{}^3\Sigma_g^+(v=4,N=10)$ fine-structure levels is attributed to tunneling predissociation through a barrier in the potential of the $c$ state and is quantitatively accounted for by calculations of the predissociation widths.
\end{abstract}

\maketitle

\section{Introduction}\label{sec:introduction}
Shortly after the initial observation of a band spectrum of helium \cite{goldstein13a,curtis13a}, Fowler \cite{fowler15a} and later Curtis and Long \cite{curtis25a} identified additional bands and realized that they could be described using the Rydberg formula, which had previously been applied exclusively to atomic spectra. This development led to significant interest in the helium bands (see Ref.~\citenum{curtis29a} for an extensive early list) making the band spectrum of helium one of the best known at that time \cite{weizel29a}. By comparing the Doppler widths of the line and band spectra, Leo \cite{leo26a} confirmed that the helium dimer (He$_2$) was the source of these bands. Weizel and F{\"u}chtbauer were the first to detect vibrationally excited states of He$_2$ \cite{weizel27a}. To explain the Rydberg-series structure of the bands, Takahashi \cite{takahashi22a} and Weizel \cite{weizel29a,weizel29b} proposed a model in which an excited electron orbits a He$_2$$^+$-ion core and spectral lines correspond to transitions between these orbits. Weizel \cite{weizel29a} even used the vibrational frequency of the low-lying terms to predict the stability and vibrational frequency of the He$_2$$^+$ ion by extrapolation of the observed series. Today, He$_2$ represents the best known example of a ``Rydberg molecule'', i.e., a molecule with a repulsive ground state and stable Rydberg states \cite{herzberg86a,herzberg87a}. In this type of molecules, the electronic states have nearly parallel potential-energy curves, which strongly disfavors electronic transitions that involve changes in vibrational excitation ($\Delta v\neq0$). Theoretical studies on the Rydberg states of diatomic molecules \cite{mulliken64b,mulliken66a,mulliken69a} and the need to better characterize the vibrational structure in the He$_2$ band spectra motivated a detailed systematic reexamination of the He$_2$ spectrum by Ginter and coworkers \cite{ginter65a,ginter65b,ginter65c,ginter66a,ginter68a,ginter70a,brown71a,brown73a,orth76a,orth77a,orth78a,ginter80a,ginter83a,ginter84a,ginter88a,ginter89a} fifty years after its discovery.

In recent years, highly accurate \textit{ab-initio} calculations including relativistic and quantum-electrodynamics (QED) effects have become feasible in molecular three- and four-electron systems \cite{yarkony89a,przybytek10a,przybytek17a,matyus18a,ferenc20a,racsai24a,jeszenszki25a,margocsy25a,racsai25a}. To test the current level of theory, precise and accurate experimental data are needed. The lowest-lying triplet states $a\,{}^3\Sigma_u^+$, $b\,{}^3\Pi_g$ and $c\,{}^3\Sigma_g^+$ of He$_2$ are ideal systems for this purpose because they are subject to a wide range of physical phenomena, including barrier tunneling \cite{lorents89a}, nonadiabatic effects \cite{ginter65c,hazell95a,li11a}, and relativistic effects such as spin-spin and spin-rotation interactions \cite{lichten74a,beck74a,bjerre98a,minaev03a}. The theoretical treatment of the fine structure of the triplet states of He$_2$ remains limited to the lowest members of the $n\mathrm{s}$ and $n\mathrm{p}$ Rydberg series \cite{beck74a,bjerre98a,minaev03a,racsai25a,margocsy25a}.

\begin{figure*}[t!]
    \centering
    \includegraphics[width=0.7\textwidth]{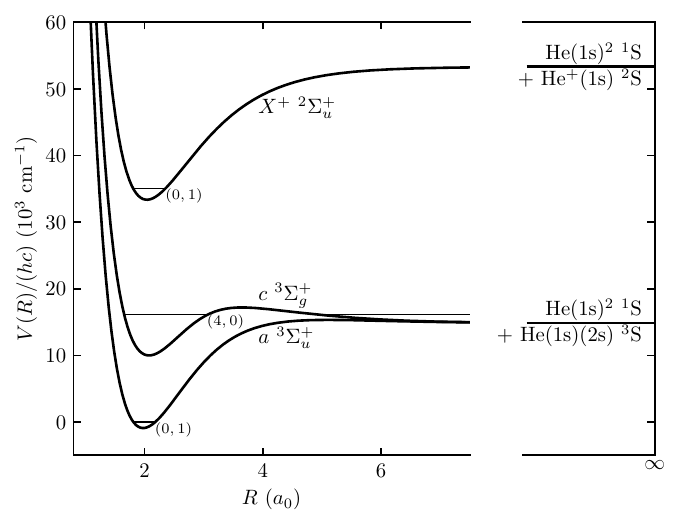}
    \caption{\label{Fig_potentials}\small Potential-energy functions of the $a\ ^3\Sigma_u^+$ and $c\ ^3\Sigma_g^+$ states of He$_2$ taken from Ref.~\cite{yarkony89a} and of the $X^+\ ^2\Sigma_u^+$ state of He$_2$$^+$ taken from Ref.~\cite{tung12a}. The energies of selected rovibrational levels $(v,N)$ are indicated as horizontal lines.}
\end{figure*}

The potential-energy function of the $X\;{}^1\Sigma_g^+$ ground state of He$_2$ is repulsive at short range and exhibits a well at long range with a single very weakly bound halo state \cite{grisenti00a,jeziorska07a,przybytek10a,zeller16a,przybytek17a}. Consequently, the radiative decay of the lowest-lying excited singlet state ($A\;{}^1\Sigma_u^+$) leads to an intense emission continuum in the extreme-ultraviolet region \cite{hopfield30b,hopfield30c,hill89a}. In contrast, the lowest-lying triplet state $a\;{}^3\Sigma_u^+$, which correlates to the $\mathrm{He(1\;{}^1S_0) + He(2\;{}^3S_1)}$ dissociation limit,
as shown in Fig.~\ref{Fig_potentials}, is metastable with a lifetime of several seconds \cite{chabalowski89a}. This state, which owes its long lifetime to the extreme weakness of the spin-orbit coupling to ${}^1\Pi_u$ states, is commonly used as an effective ground state in spectroscopic studies of He$_2$. Whereas the potential-energy curve of the $a\;{}^3\Sigma_u^+$ state closely resembles that of the ionic $X^+\;{}^2\Sigma_u^+$ ground state (see Fig.~\ref{Fig_potentials}) -- as expected for a Rydberg molecule -- the potential-energy curve of the $c\;{}^3\Sigma_g^+$ state strongly deviates from that of the ionic ground state at large internuclear distances $R$ and exhibits a potential hump with a local maximum near $R = 3.7\,a_0$ and $E/(hc)=2400\,\mathrm{cm}^{-1}$ above the dissociation limit \cite{jordan86a,lorents89a}. This hump results from an avoided crossing between an attractive Rydberg state with a $(1\mathrm{s}\sigma_g)^2(1\mathrm{s}\sigma_u)^1$ ion-core and a repulsive Rydberg state with $(1\mathrm{s}\sigma_g)^1(1\mathrm{s}\sigma_u)^2$ ion-core configuration \cite{guberman75a}, and leads to competing decay channels for all rovibrational levels located energetically above the dissociation limit. These levels are quasibound shape resonances and can either fluoresce down to the $a\;{}^3\Sigma_u^+$ state or tunnel through the barrier, leading to the $\mathrm{He(1\;^1S_0) + He(2\;^3S_1})$ dissociation products \cite{lorents89a,hazell95a,roney97a}. All rotational levels of the $c\;{}^3\Sigma_g^+(v=4)$ state lie above the dissociation limit, making the $c\;{}^3\Sigma_g^+(v=4)$ state an ideal candidate to study these complex decay dynamics. So far, the $c\;{}^3\Sigma_g^+ - a\;{}^3\Sigma_u^+$ system of He$_2$ has only been investigated in a few experiments \cite{meggers32a,cuthbertson53a,ginter65a,lorents89a,kristensen90a,hazell95a,li10a,focsa98a}.

\begin{figure*}[t!]
    \centering
    \includegraphics[width=\textwidth]{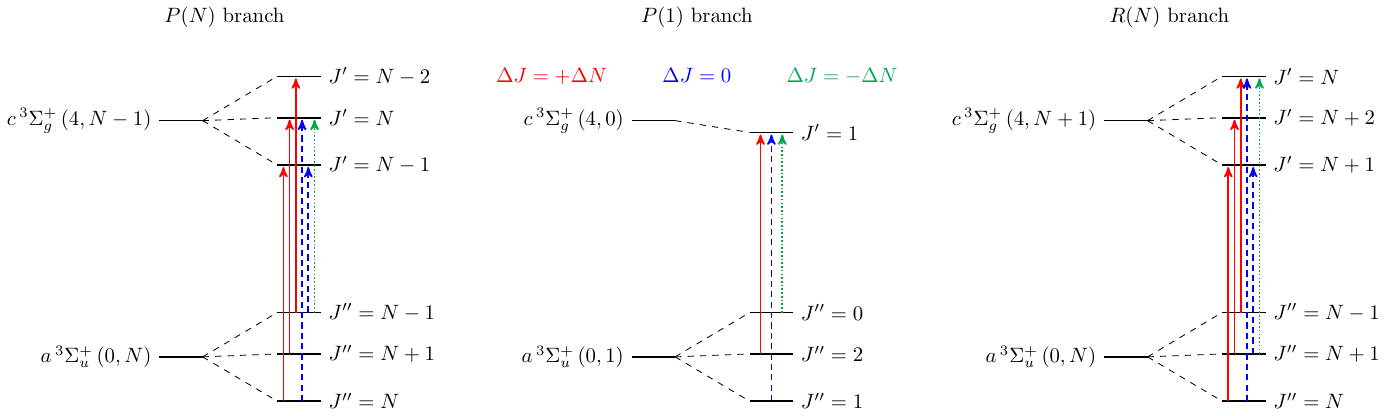}
    \caption{\label{fig:FS_transitions}\small Dipole-allowed transitions between the fine-structure levels of the $a\;{}^3\Sigma_u^+(v=0)$ and $c\;{}^3\Sigma_g^+(v=4)$ state. The solid red, dashed blue and dotted green arrows designate transitions with $\Delta J = +\Delta N$, $\Delta J = 0$ and $\Delta J = -\Delta N$, respectively.}
\end{figure*}

Because in He$_2$ the spin-orbit and spin-spin splittings are much smaller than the spacings between adjacent rotational level , the $a$ and $c$ states are well described in Hund's angular-momentum-coupling case (b) \cite{lefebvrebrion04a}. In ${}^3\Sigma$ states, each rotational level with rotational-angular-momentum quantum number $N$ is split by the spin-spin and spin-rotation interactions into three fine-structure levels with total-angular-momentum quantum number $J = N,N\pm1$ \cite{kramers29a,kramers29b,lefebvrebrion04a}. Only even (odd) values of $N$ have nonzero spin-statistical weights in electronic states of $\Sigma_g^+$ ($\Sigma_u^+$) symmetry of ${}^4$He$_2$. The dominant electric-dipole allowed transitions in the ${c\;{}^3\Sigma_g^+-a\;{}^3\Sigma_u^+}$ band system are shown in Fig.~\ref{fig:FS_transitions} and obey the selection rules $\Delta J=0,\pm1$ and $\Delta N=\pm1$. In principle, transitions with $\Delta N=\pm3$ and $\Delta J=\pm1$ are also allowed in the electric-dipole approximation, as $N$ is not a good quantum number because of the spin-spin interaction. However, the $\Delta N = \pm2$ coupling of adjacent rotational levels is extremely weak and $\Delta N = \pm3$ transitions are hardly observable.

Lichten and coworkers have measured the fine-structure splittings of the $a\;{}^3\Sigma_u^+(v=0)$ state by molecular-beam magnetic-resonance techniques \cite{lichten74a,vierima75a,lichten78a}, achieving a precision of around 10\,kHz. Reaching a comparable precision, Bjerre and collaborators employed a laser--radio-frequency (rf) double-resonance method to probe the fine-structure splittings in higher rotational levels of the $a\;{}^3\Sigma_u^+({v=0},{N=7,9,11,25,27,29})$ and $a\;{}^3\Sigma_u^+({v=2},{N=7,9,11})$ states \cite{lorents89a,kristensen90a,hazell95a}. Rogers et al. \cite{rogers88a} resolved individual transitions between fine-structure levels of the $0\rightarrow0$ and $1\rightarrow1$ vibrational bands of the $b\;{}^3\Pi_g-a\;{}^3\Sigma_u^+$ system using Fourier-transform emission spectroscopy and Focsa et al. \cite{focsa98a} carried out a global analysis of the six lowest-lying electronic states of He$_2$, including these data as well as three additional fine-structure intervals for the $a\;{}^3\Sigma_u^+$ state ($v=1,N=25,27$) and ($v=2,N=19$), as communicated by Bjerre \cite{focsa98a}. Recently, Semeria et al. \cite{semeria18a} performed rf measurements using Zeeman-decelerated and spin-polarized molecular beams to determine the fine-structure splittings of the lowest eleven rotational levels of the $a\;{}^3\Sigma_u^+(v=0)$ and the lowest two rotational levels of the $v=1$ state, reaching a precision of 290\,Hz for the rovibrational ground state.

In contrast, the fine structure of the $c\;{}^3\Sigma_g^+$ state remains poorly characterized. It has only been determined for two rotational levels $({N = 26, 28})$ of the ${v=2}$ state in laser-rf double-resonance measurements and for three rotational levels $(N = 8, 10, 12)$ of the ${v=4}$ state by fast-beam photofragment spectroscopy in pioneering measurements by Bjerre and coworkers \cite{hazell95a,kristensen90a}. However, the resolution of $\sim80\,\mathrm{MHz}$ for the measurements of the ${v=4}$ state was insufficient to fully resolve the fine structure.

We report here on a precision spectroscopic study of optical transitions between the fine-structure levels of the ${a\;{}^3\Sigma_u^+(v=0,N=1-9)}$ and ${c\;{}^3\Sigma_g^+(v=4,N=0-10)}$ states of He$_2$. A resonance-enhanced two-photon ionization scheme with two perpendicularly crossed laser beams enabled sub-Doppler measurements. Using a calibration procedure based on an optical frequency comb and the retroreflection of the laser beam to cancel first-order Doppler shifts, absolute transition frequencies could be determined with a relative accuracy ($\Delta\nu/\nu$) of $2.5\times10^{-10}$ and the fine-structure of the $c\;{}^3\Sigma_g^+(v=4)-a\;{}^3\Sigma_u^+(v=0)$ band could be fully resolved in most rotational branches. By combining these new measurements with earlier data from Refs.~\citenum{kristensen90a,semeria18a,hazell95a}, the molecular constants of the $a\;{}^3\Sigma_u^+(v=0)$ and $c\;{}^3\Sigma_g^+(v=4)$ states of ${}^4$He$_2$ could be significantly improved in a global nonlinear least-squares fit. Additionally, the state-dependent lifetimes of the $c\;{}^3\Sigma_g^+(v=4,N)$ fine-structure levels caused by competing deexcitation channels (radiative decay and predissociation) were investigated, both experimentally and computationally. The new results served as reference data to benchmark a new generation of highly accurate calculations carried out in a parallel effort by M{\'a}tyus and coworkers \cite{jeszenszki25a,margocsy25a,racsai25a}.

\section{Experimental}\label{sec:experimental}
\begin{figure*}[t!]
    \centering
    \includegraphics[width=\textwidth]{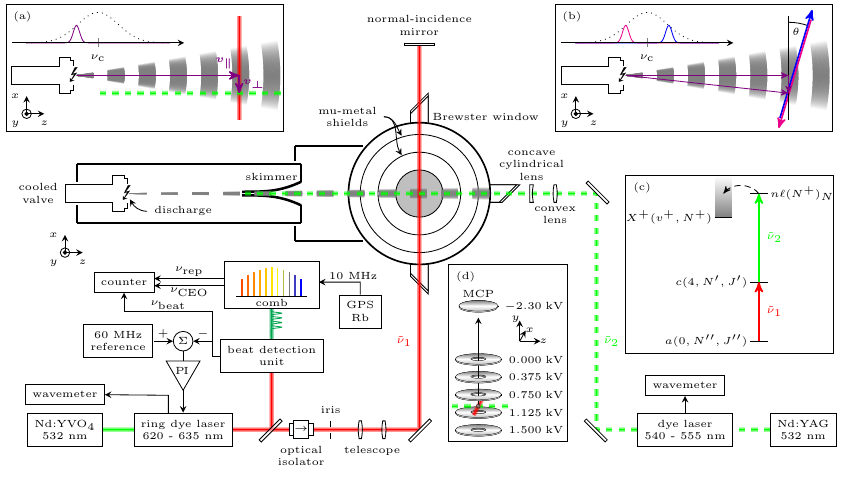}
    \caption{\label{fig:setup}\small Schematic representation of the experimental setup. The insets illustrate the methods used to reduce the Doppler width (a) and correct the Doppler shift (b), the excitation and detection scheme (c), and the time-of-flight spectrometer (d). The solid red and dashed green lines represent the beams of lasers 1 and 2, respectively, which cross at right angles in the photoexcitation region.}
\end{figure*}

The experimental setup is shown schematically in Fig.~\ref{fig:setup}. Metastable He$_2$ molecules in the $a\;{}^3\Sigma_u^+$ state are produced in a supersonic expansion of pure helium gas using a dielectric-barrier discharge at the orifice of a pulsed valve. The body of the valve is cooled to 77\,K, resulting in a forward velocity of the molecular beam of approximately $\mathrm{1000\,m\,s^{-1}}$. The expansion angle of the molecular beam is reduced using a 3-mm-diameter skimmer placed 40\,cm downstream from the valve, which also allows efficient differential pumping with a time-averaged background pressure of only $\sim5\times10^{-7}\,\mathrm{mbar}$ in the photoexcitation region.

The $c\;{}^3\Sigma_g^+-a\;{}^3\Sigma_u^+$ transitions were measured using the ($1+1^\prime$) Resonance-Enhanced-Two-Photon-Ionization (R2PI) scheme depicted in Fig.~\ref{fig:setup}(c). A narrow-band continuous-wave (cw) laser (wavenumber $\tilde{\nu}_1$) was used to resolve the fine-structure of the $c\;{}^3\Sigma_g^+-a\;{}^3\Sigma_u^+$ transitions and a broadband pulsed dye laser ($\tilde{\nu}_2$) to selectively excite the $c\;{}^3\Sigma_g^+(v=4)$ fine-structure levels to autoionizing Rydberg states ($n\ell(N^+)_N$) located above the $X^+\;{}^2\Sigma_u^+(v^+=0,N^+=1)$ ionization threshold for efficient detection. The resulting He$_2$$^+$ ions were extracted 3\,$\mathrm{\mu s}$ after the laser pulse ($\tilde{\nu}_2$) by applying a pulsed electric potential of 1.5\,kV across five resistively coupled parallel electrodes for detection at a microchannel-plate (MCP) detector, as schematically represented in Fig.~\ref{fig:setup}(d). The integrated He$_2$$^+$-ion signal was recorded as a function of the wavenumber $\tilde{\nu}_1$ of the first laser.

The narrow-bandwidth cw radiation was generated using a ring dye laser (Sirah Matisse 2 DS) locked to a frequency comb (Menlo Systems Inc., FC1500-250-WG) referenced to a GPS-disciplined rubidium frequency standard (Stanford Research Systems FS725; GPS receiver: Stanford Research Systems, FS752) \cite{hoelsch19a}. The stabilization of the ring-dye-laser frequency was achieved by locking the beat signal ($\nu_\mathrm{beat}$) between the ring-dye laser and a tooth of the frequency-doubled output of the frequency comb to a 60\,MHz reference using a frequency discriminator \cite{ritt04a,herburger24a}. The generated error signal from the frequency discriminator was used to adjust the reference cavity of the ring dye laser through a proportional-integral (PI) loop. The frequency $\nu_L$ of the laser was determined according to $\nu_L = n\nu_\mathrm{rep}\pm 2\nu_\mathrm{CEO}\pm\nu_\mathrm{beat}$, where $\nu_\mathrm{rep}\approx250\,\mathrm{MHz}$ is the repetition rate of the frequency comb, $\nu_\mathrm{CEO}=20\,\mathrm{MHz}$ its carrier-envelope-offset frequency and $n$ the comb-mode index, which was determined with a wavemeter (HighFinesse WS7-60) specified to an absolute accuracy of 60\,MHz ($3\sigma$). The frequency of the ring-dye laser was scanned by varying the repetition rate of the frequency comb.

To reduce the Doppler broadening caused by the narrow expansion angle of the supersonic beam, the cw and pulsed laser beams were aligned perpendicularly in the $xz$-plane (see Fig.~\ref{fig:setup}). The cw laser crossed the molecular beam at near right angles, and the pulsed laser propagated antiparallel to the molecular beam. The pulsed laser beam was focused along the $x$-axis and expanded along the $y$-axis using a convex lens and a concave cylindrical lens, respectively. This arrangement resulted in molecules within a narrow transverse-velocity group being ionized and detected \cite{clausen23a}, as illustrated in Fig.~\ref{fig:setup}(a). However, even for a perfect perpendicular arrangement of the propagation axes of cw-laser and molecular beams, a systematic Doppler shift $-\vec{k}\cdot\vec{v}_\perp$ is introduced if the intersection point of the lasers is not located exactly on the molecular-beam axis, where $\vec{k}$ is the wavevector of the cw laser and $\vec{v}_\perp$ the transverse velocity of the molecules. Moreover, an additional Doppler shift $-\vec{k}\cdot\vec{v}_\parallel$ results from the high forward velocity $\vec{v}_\parallel\approx1000\,\mathrm{m\,s^{-1}}$ of the molecular beam whenever the angle between the propagation axes of the molecular and cw-laser beams deviate from 90$^{\circ}$ by a misalignment angle $\theta$. To compensate for these systematic shifts, the cw-laser beam was weakly focused with a telescope onto a normal-incidence mirror and retroreflected. A high degree of collinearity, with a maximum deviation angle of $\theta=0.15\,\mathrm{mrad}$, was achieved by aligning the retroreflected beam through an iris positioned 3.4\,m away from the normal-incidence mirror. This configuration splits each line into a red- and a blue-shifted component  ($\nu_\mathrm{r}$ and $\nu_\mathrm{b}$, respectively). The first-order Doppler-corrected frequency $\nu_\mathrm{c}$ is then given by $(\nu_\mathrm{r}+\nu_\mathrm{b})/2$, as illustrated in Fig.~\ref{fig:setup}(b).

For the lifetime measurements discussed in Section \ref{sec:decay_dynamics}, the reflected laser beam was blocked and a pulsed acousto-optic modulator (Brimrose GPF-1000-500-800) operated at a modulation frequency of $\nu_\mathrm{AOM}=+1\,\mathrm{GHz}$ was used to cut 3-$\mu\mathrm{s}$-long pulses out of the cw-laser beam. The first-order diffraction output of the AOM was used to drive selected transitions between the fine-structure levels of the $a\;{}^3\Sigma_u^+$ and $c\;{}^3\Sigma_g^+$ states. A delay generator (Stanford Research Systems, Model DG535) was used to adjust the delay time between the falling edge of the cw-laser pulse and the dye-laser pulse used to ionize the $c$-state levels. Monitoring the He$_2$$^+$ signal as a function of this delay time led to the observation of an exponential decay caused by the joined effects of the fluorescence and tunneling predissociation of the $c$-state levels.

\section{Results and Discussion}\label{sec:results}
\subsection{Overall Structure of the Spectra and Uncertainties}\label{sec:spectroscopic_results}

\begin{figure*}
    \includegraphics[width=0.84\linewidth]{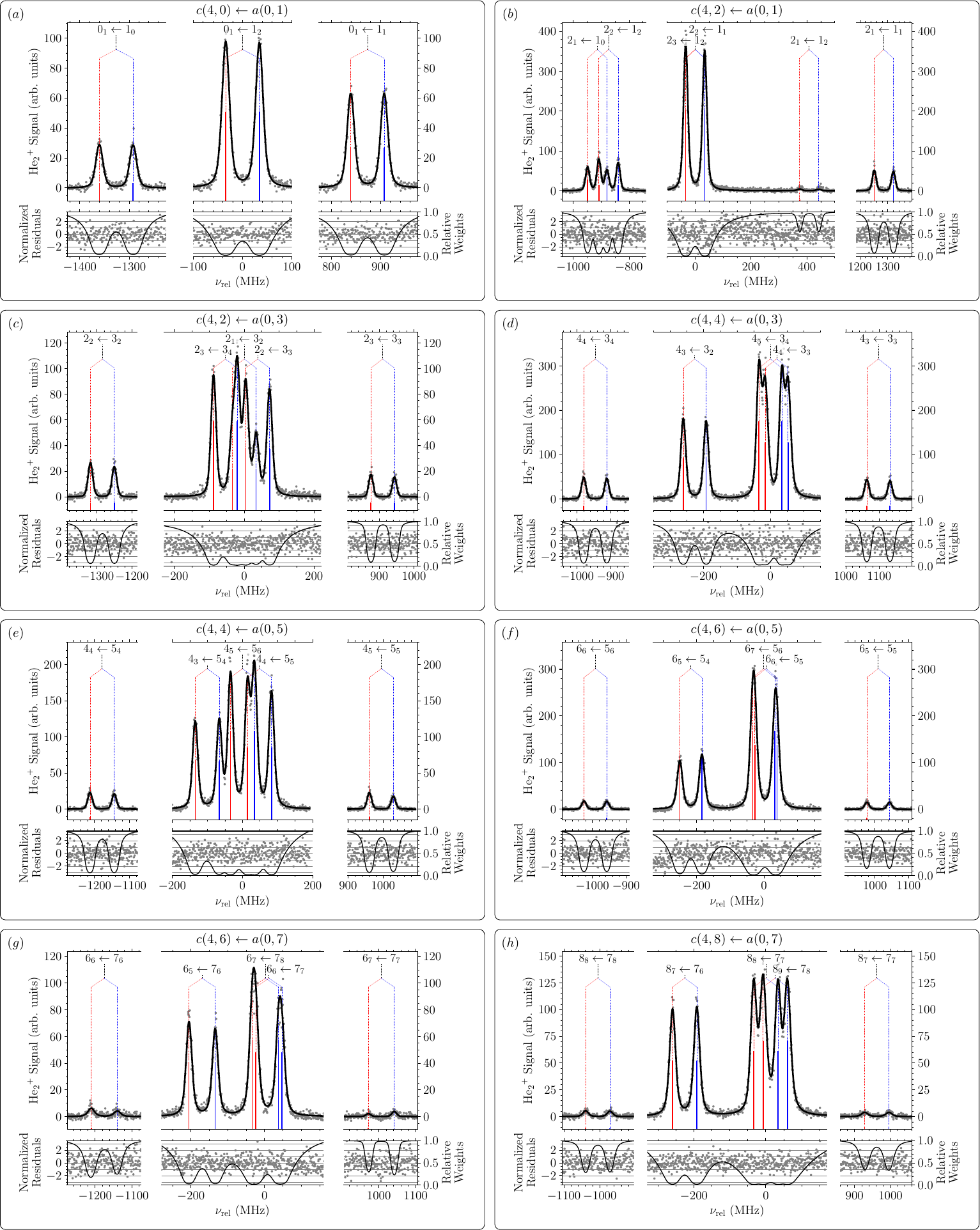}
    \caption{\label{fig:RP_1to7}\small Fine-structure-resolved spectra of the $c\;{}^3\Sigma_g^+({v^{\prime}=4},N^{\prime})\leftarrow a\;{}^3\Sigma_u^+({v^{\prime\prime}=0},N^{\prime\prime})$ transitions of ${}^4$He$_2$. The red and blue sticks indicate the fitted central positions of the two Doppler components. Their relative strength has been scaled according to Eq.~(\ref{eq:rel_intensities}). The assignments are given as $N^{\prime}_{J^{\prime}} \leftarrow N^{\prime\prime}_{J^{\prime\prime}}$. In the upper plots of each panel, the grey dots and full black lines represent the measured signal intensities and fitted intensity distributions, respectively. The grey dots and full black lines in the lower plots represent the normalized residuals and the corresponding weights of the data points used in the least-squares fit, respectively. See text for details.}
\end{figure*}

\begin{figure*}[t]
    \includegraphics[width=\linewidth]{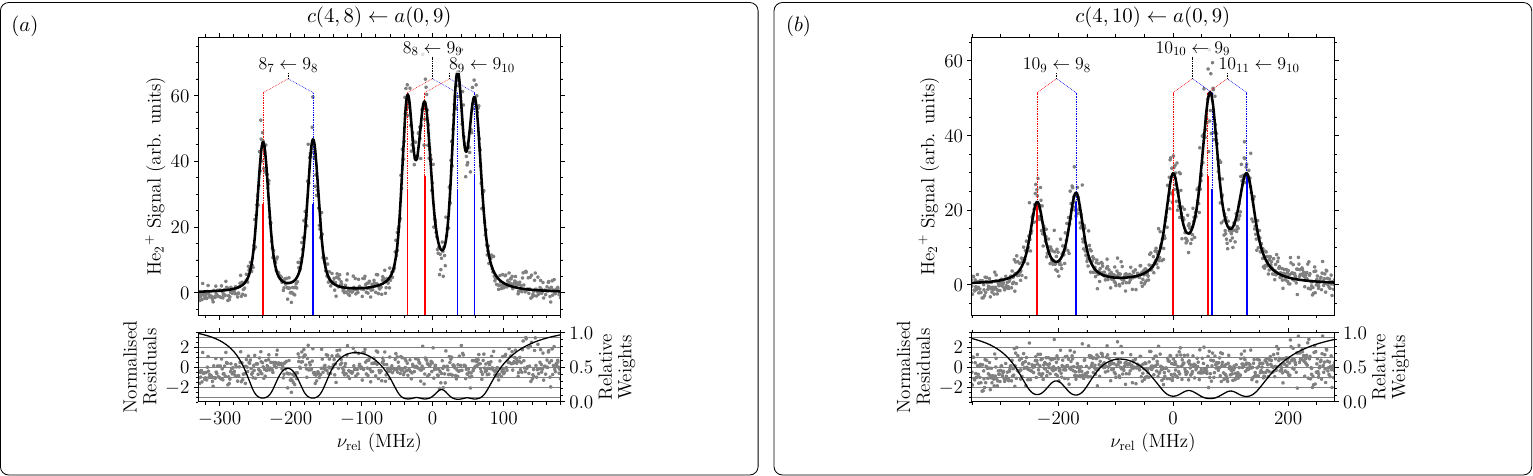}
    \caption{\label{fig:RP_9}\small Fine-structure-resolved spectra of the $c\;{}^3\Sigma_g^+({v^{\prime}=4},N^{\prime})\leftarrow a\;{}^3\Sigma_u^+({v^{\prime\prime}=0},N^{\prime\prime})$ transitions of ${}^4$He$_2$. The red and blue sticks indicate the fitted central positions of the two Doppler components. Their relative strength has been scaled according to Eq.~(\ref{eq:rel_intensities}). The assignments are given as $N^{\prime}_{J^{\prime}} \leftarrow N^{\prime\prime}_{J^{\prime\prime}}$. In the upper plots of each panel, the grey dots and full black lines represent the measured signal intensities and fitted intensity distributions, respectively. The grey dots and full black lines in the lower plots represent the normalized residuals and the corresponding weights of the data points used in the least-squares fit, respectively. See text for details.}
\end{figure*}

High-resolution spectra of the transitions between the fine-structure levels of the $a\;{}^3\Sigma_u^+(v^{\prime\prime}=0,N^{\prime\prime})$ and $c\;{}^3\Sigma_g^+({v^{\prime}=4},N^{\prime})$ states with $N^{\prime\prime}\in[1,3,5,7,9]$ and $N^{\prime}\in[0, 2, 4, 6, 8, 10]$ measured in the present work are presented in Figs.~\ref{fig:RP_1to7} and \ref{fig:RP_9}. 
As explained in Section~\ref{sec:experimental}, each transition is split into a red- and a blue-shifted Doppler component. These components are indicated by red and blue sticks, respectively, and are separated by a splitting of $2\Delta_\mathrm{D}\approx68\,\mathrm{MHz}$, corresponding to an angle of $\theta\approx1.1^\circ$ [see Fig.~\ref{fig:setup}(b)] for $v_\parallel=1000\,\mathrm{m\,s^{-1}}$ and $v_\perp=2\,\mathrm{m\,s^{-1}}$ ($\Delta_\mathrm{D}=\frac{\nu}{c}[\sin\theta v_\parallel + \cos\theta v_\perp])$. Because of the narrow full width at half maximum (FWHM) of $\varGamma\approx15\,\mathrm{MHz}$ of the spectral lines, the two Doppler components are clearly separated, enabling the accurate determination of the first-order Doppler-corrected frequency $\nu_\mathrm{c}$.

Except for the $c(4,0) \leftarrow a(0,1)$ and $c(4,2) \leftarrow a(0,1)$ transitions [see Figs.~\ref{fig:RP_1to7}(a) and (b)], all transitions exhibit a similar pattern. They consist of three strong central components, with $\Delta J=\Delta N$ (red lines in Fig.~\ref{fig:FS_transitions}), surrounded by two satellites shifted by approximately 1\,GHz to the red and the blue with ${\Delta J=0}$ (blue dashed lines in Fig.~\ref{fig:FS_transitions}). The line with $\Delta J=-\Delta N$ (green dotted line in Fig.~\ref{fig:FS_transitions}) is in general too weak to be observed, except at the lowest $N$ values (see below). Considering the general level structure depicted in Fig.~\ref{fig:FS_transitions}, the observed pattern suggests that the fine-structure splittings of the $a\;{}^3\Sigma_u^+(v=0)$ and $c\;{}^3\Sigma_g^+(v=4)$ states are similar. A closer inspection of the spectral patterns reveals that the three central components, $\nu_{-}=\nu(J^{\prime}=N^{\prime}-1)$, $\nu_{=}=\nu(J^{\prime}=N^{\prime})$ and $\nu_{+}=\nu(J^{\prime}=N^{\prime}+1)$, reverse their order from $\nu_{-}<\nu_{+}<\nu_{=}$ for $4\leq{N^{\prime}\leq6}$ to ${\nu_{-}<\nu_{=}<\nu_{+}}$ for $N^{\prime} \geq 8$. This change in the order leads to the conclusion that the fine-structure intervals of the $c\;{}^3\Sigma_g^+(v=4)$ and $a\;{}^3\Sigma_u^+(v=0)$ states evolve differently as $N$ increases.

The $c(4,0) \leftarrow a(0,1)$ and $c(4,2) \leftarrow a(0,1)$ transitions [see Figs.~\ref{fig:RP_1to7}(a) and (b)] exhibit qualitatively different patterns. In the case of the $c(4,0) \leftarrow a(0,1)$ transitions, only one upper fine-structure level with $N=0$ and $J=1$ exists, resulting in only three instead of six possible transitions (see Fig.~\ref{fig:FS_transitions}). Consequently, the splittings of these three transitions directly reflect the fine-structure intervals of the $a\;{}^3\Sigma_u^+({v=0,N=1})$ state. In the case of the $c(4,2) \leftarrow a(0,1)$ transitions, the $2_1\leftarrow1_0$ transition is shifted to lower frequencies by approximately 0.9\,GHz compared to the usual pattern of three central components, and is even located below the weak red-shifted $\Delta J = 0$ ($2_2\leftarrow1_2$) satellite (green dashed line in Fig.~\ref{fig:FS_transitions}). This situation results from the fact that the interval between the $1_0$ and $1_1$ fine-structure components of the $a\;{}^3\Sigma_u^+({v=0,N=1})$ level is approximately twice as large as all other $N_{N-1} - N_N$ intervals of the $a\;{}^3\Sigma_u^+(v=0)$ state \cite{semeria18a}.

In the electric-dipole approximation, the relative intensities of the transitions between two fine-structure components of any $N^{\prime} \leftarrow N^{\prime\prime}$ transition are given by \cite{verdegay25a}
\begin{equation}\label{eq:rel_intensities}
    \mathcal{I} = (2J^{\prime\prime}+1)(2J^{\prime}+1)
    \left\{\begin{array}{ccc}
    N^{\prime\prime} & J^{\prime\prime} & S \\
    J^{\prime} & N^{\prime} & 1
    \end{array}\right\}^2,
\end{equation}
where $\{:::\}$ represents a Wigner-$6j$ symbol. The measured intensities $\mathcal{I}(\Delta J)$ thus follow the order $\mathcal{I}({\Delta J = \Delta N}) > \mathcal{I}({\Delta J = 0}) \gg \mathcal{I}({\Delta J = -\Delta N})$, and this trend becomes more pronounced with increasing $N$ value. In Fig.~\ref{fig:RP_1to7}(a) and (b) the weak ${\Delta J = -\Delta N}$ transitions, which had not been detected in previous work, are clearly identified for the $c(4,0) \leftarrow a(0,1)$ and $c(4,2) \leftarrow a(0,1)$ transitions. Overall the relative intensities in Fig.~\ref{fig:RP_1to7} follow the relative trends predicted by Eq.~(\ref{eq:rel_intensities}), with one noticeable exception: the relative intensities of the $2_2\leftarrow1_2$ and $2_1\leftarrow1_0$ transitions in the low-frequency line cluster [see Fig.~\ref{fig:RP_1to7}(b)] are opposite to the calculated ones (indicated by the magnitude of the sticks). An incorrect assignment of the transitions can be ruled out by forming the combination difference of the $2_1\leftarrow1_0$ with the $2_1\leftarrow1_1$ transition, which corresponds to the $1_0 - 1_1$ splitting of the $a\;{}^3\Sigma_u^+(v=0)$ state. This splitting has been measured with high accuracy to be 2199.971\,47(29)\,MHz \cite{semeria18a}. For the assignment given in Fig.~\ref{fig:RP_1to7}(b), the combination difference yields a value of 2199.5(6)\,MHz for this splitting, while reversing the assignment of the $2_2\leftarrow1_2$ and $2_1\leftarrow1_0$ lines would yield a value of 2159.3(6)\,MHz, which is 40\,MHz too low. The reversed intensity pattern might be the result of the finite bandwidth of the pulsed laser used in the ionization step. For all transitions to a particular upper rotational level $N^\prime$, the pulsed-laser frequency was kept constant. For large fine-structure intervals of a particular $N^\prime$ level, the position and intensity distribution of the pulsed laser might have led to different ionization efficiencies of the fine-structure components. Alternatively, the intensity reversal might originate from intensity borrowing to $b\;{}^3\Pi_g\leftarrow a\;{}^3\Sigma_u^+$ transitions induced by spin-orbit and nonadiabatic couplings between the $c\;{}^3\Sigma_g^+$ and $b\;{}^3\Pi_g$ states.

Because of the decreasing intensities of the ${\Delta J = 0}$ transitions with increasing $N$, only the central ${\Delta J = \Delta N}$ components could be observed for the $c(4,8) \leftarrow a(0,9)$ and $c(4,10) \leftarrow a(0,9)$ transitions, as seen in Fig.~\ref{fig:RP_9}. Moreover, the lines of the $c(4,10) \leftarrow a(0,9)$ transitions are significantly broader than the other lines of the $c-a$ band system observed in this work. The origin of this broadening is discussed in Section~\ref{sec:decay_dynamics}.

\begin{table}[t!]
    \caption{\label{tab:uncertainties}\small Overview of systematic shifts and uncertainties. All values are in kHz.}
    \begin{ruledtabular}
    \begin{tabular}{l D{.}{}{3.1} D{.}{}{3.0}}
        Source & \multicolumn{1}{c}{\textrm{Shift}} & \multicolumn{1}{c}{\textrm{Uncertainty}} \\
        \colrule
        Residual first-order Doppler shift  &                                                                 & 110. \\
        Second-order Doppler shift          &  -3.\footnote{Exact value depends on the transition frequency.} & \\
        Photon-recoil shift                 & +64.\footnotemark[1]                                            & \\
        ac Stark shift                      &                                                                 &  90. \\
        Frequency Calibration               &                                                                 & <10. \\
        Zeeman shift                        &                                                                 &  <1. \\
        \colrule
        $\sigma_\mathrm{sys}$               &                                                                 & 200. \\
    \end{tabular}
    \end{ruledtabular}
\end{table}

Systematic uncertainties and shifts for the measured transition frequencies are summarized in Table~\ref{tab:uncertainties}. The main source of systematic uncertainty is caused by the residual first-order Doppler shift resulting from the retroreflected and incoming cw-laser-beam axes not being exactly antiparallel. The uncertainty was estimated from multiple measurements of all fine-structure components of the $c(4,0) \leftarrow a(0,1)$ transitions on three different measurement days and after full realignment of the optical setup. In addition, for the $0_1\leftarrow1_2$ transition, the retroreflection was intentionally adjusted to the edge of the alignment iris before recording a spectrum. From the resulting Doppler shift, the systematic uncertainty was estimated to be 110\,kHz. 

The second-order Doppler effect leads to a systematic red shift $\Delta\nu = -\nu v_\mathrm{beam}^2/(2c^2)$. At the mean velocity $v_\mathrm{beam}=1000\,\mathrm{m\,s^{-1}}$ of the molecular beam, this shift is around 2.7\,kHz. The photon recoil shift $\Delta\nu_\mathrm{recoil} = h\nu^2/(2m_\mathrm{He_2}c^2)$ results in blue-shifted transition frequencies. The exact values of both shifts were calculated and subtracted from each measured transition frequency. To assess the ac Stark shift, the three components of the $c(4,0) \leftarrow a(0,1)$ transition were measured at cw-laser powers of 50\,mW and 15\,mW. Because the transition frequencies measured at these powers agreed within the statistical uncertainty of 90\,kHz, the ac Stark shift was deemed to be insignificant for the typical laser powers used in the experiments, and a conservative systematic uncertainty of 90\,kHz was assumed. The accuracy of the frequency calibration is limited by the stability of the GPS-disciplined Rb clock, resulting in a systematic uncertainty of $<\mathrm{10\,kHz}$ \cite{clausen21a}. The fine-structure levels of He$_2$ are subject to Zeeman shifts. The two concentric mu-metal shields reduce the magnetic field inside the excitation region to values of less than 10\,mG. However, even at 0.5\,G, the centroid positions of the $m_J$ manifolds would not shift by more than 1\,kHz from their field-free position. Consequently, Zeeman shifts are negligible in our experiment.

An overall systematic uncertainty of 200\,kHz corresponding to the sum of individual contributions in Table~\ref{tab:uncertainties} was used for all transitions except the $c(4,0) \leftarrow a(0,1)$ transitions, for which the statistical uncertainty of the independent measurements was taken. 

\subsection{Line Shapes and Transition Frequencies}\label{sec:fitting}
The transition frequencies were determined in a fit procedure relying on the fact that the lineshapes are described by Voigt profiles. The Gaussian part, with width $\varGamma_\mathrm{G}$, incorporates the residual Doppler broadening originating from the finite size of the pulsed laser in the direction of the cw laser, i.e., along the $x$-axis in Fig.~\ref{fig:setup}(a). The Lorentzian part $\varGamma_\mathrm{L}$ includes effects from the natural linewidth of the transitions and power broadening. The analysis of the three fully resolved components of the $c(4,0) \leftarrow a(0,1)$ transition [see Fig.~\ref{fig:RP_1to7}(a)] led to the conclusion that $\varGamma_\mathrm{G} \approx 13.5\,\mathrm{MHz}$, which was then kept fixed in the line-shape fits of all other transitions, after we verified that changing $\varGamma_\mathrm{G}$ between 12.5\,MHz and 14.5\,MHz did not significantly change the central positions of the fitted lines. 

\begin{table*}[p]
    \renewcommand{\arraystretch}{0.75}
    \caption{\label{tab:transitions}\small Fitted central positions $\tilde{\nu}$ of the transitions between the fine-structure levels of the $c\;{}^3\Sigma_g^+(v=4)$ and $a\;{}^3\Sigma_u^+(v=0)$ states of ${}^4$He$_2$. The numbers in parentheses are the statistical uncertainties in units of the last digit. All transitions with $N^\prime>0$ have an additional systematic uncertainty of 200\,kHz ($6.7\times10^{-6}\,\mathrm{cm^{-1}}$). All frequencies were corrected for the second-order Doppler and photon-recoil shifts.}
    \begin{ruledtabular}
    \begin{tabular}{
        D{.}{\ \leftarrow\ }{4,4}
        D{.}{.}{6.11}
        p{0.05\linewidth}
        D{.}{\ \leftarrow\ }{4,4}
        D{.}{.}{6.11}
    }
    \multicolumn{2}{c}{$\Delta N = -1$} & & \multicolumn{2}{c}{$\Delta N = +1$} \\ \cline{1-2} \cline{4-5}
    N^{\prime}_{J^{\prime}}.N^{\prime\prime}_{J^{\prime\prime}} & \multicolumn{1}{c}{$\tilde{\nu}$ $(\mathrm{cm}^{-1}$)}
    & &
    N^{\prime}_{J^{\prime}}.N^{\prime\prime}_{J^{\prime\prime}} & \multicolumn{1}{c}{$\tilde{\nu}$ $(\mathrm{cm}^{-1}$)} \\
    \colrule
    0_{1}.1_{0} & 16\,069.373\,9779(62)                       & & 2_{1}.1_{0} & 16\,101.427\,3184(80)                       \\
    0_{1}.1_{2} & 16\,069.418\,2231(32)                       & & 2_{2}.1_{2} & 16\,101.428\,6604(66)                       \\
    0_{1}.1_{1} & 16\,069.447\,3586(41)                       & & 2_{3}.1_{2} & 16\,101.457\,740(25)\footnote{Unreliable fit because of overlap with other transitions.} \\
                &                                             & & 2_{2}.1_{1} & 16\,101.457\,819(42)\footnotemark[1] \\
                &                                             & & 2_{1}.1_{2} & 16\,101.471\,466(91)\footnotemark[1] \\
                &                                             & & 2_{1}.1_{1} & 16\,101.500\,6855(74)                       \\    
    \\
    2_{2}.3_{2} & 16\,025.600\,626(10)                        & & 4_{4}.3_{4} & 16\,100.096\,9135(71)                       \\
    2_{3}.3_{4} & 16\,025.641\,6899(72)                       & & 4_{3}.3_{2} & 16\,100.120\,8936(38)                       \\
    2_{1}.3_{2} & 16\,025.643\,491(15)                        & & 4_{5}.3_{4} & 16\,100.128\,4692(61)                       \\
    2_{2}.3_{3} & 16\,025.644\,776(16)                        & & 4_{4}.3_{3} & 16\,100.129\,1050(44)                       \\
    2_{3}.3_{3} & 16\,025.673\,850(12)                        & & 4_{3}.3_{3} & 16\,100.165\,0607(82)                       \\
    \\
    4_{4}.5_{4} & 15\,963.907\,097(10)                        & & 6_{6}.5_{6} & 16\,080.221\,703(13)                        \\
    4_{3}.5_{4} & 15\,963.943\,0492(54)                       & & 6_{5}.5_{4} & 16\,080.247\,5128(52)                       \\
    4_{5}.5_{6} & 15\,963.946\,4061(54)                       & & 6_{7}.5_{6} & 16\,080.254\,737(19)\footnotemark[1]                        \\
    4_{4}.5_{5} & 15\,963.948\,0207(50)                       & & 6_{6}.5_{5} & 16\,080.254\,930(22)\footnotemark[1]                        \\
    4_{5}.5_{5} & 15\,963.979\,597(11)                        & & 6_{5}.5_{5} & 16\,080.288\,424(16)                        \\
    \\
    6_{6}.7_{6} & 15\,884.148\,136(32)                        & & 8_{8}.7_{8} & 16\,041.265\,098(34)                        \\
    6_{5}.7_{6} & 15\,884.181\,6855(64)                       & & 8_{7}.7_{6} & 16\,041.291\,1125(57)                       \\
    6_{7}.7_{8} & 15\,884.187\,416(20)                        & & 8_{8}.7_{7} & 16\,041.298\,6494(96)                       \\
    6_{6}.7_{7} & 15\,884.187\,756(21)                        & & 8_{9}.7_{8} & 16\,041.299\,534(10)                        \\
    6_{7}.7_{7} & 15\,884.220\,930(43)                        & & 8_{7}.7_{7} & 16\,041.330\,737(44)                        \\
    \\
    8_{7}.9_{8} & 15\,786.030\,6834(88)                       & & 10_{9}.9_{8} & 15\,982.487\,079(23)                        \\
    8_{8}.9_{9} & 15\,786.037\,483(14)                        & & 10_{10}.9_{9} & 15\,982.494\,977(38)                        \\
    8_{9}.9_{10}& 15\,786.038\,306(17)                        & & 10_{11}.9_{10} & 15\,982.496\,992(39)                        \\
    \end{tabular}
    \end{ruledtabular}
\end{table*}

For the fitting procedure, all spectra were multiplied by the same scaling factor chosen to ensure that the normalized residuals \cite{hughes10a}
\begin{equation}\label{eq:normalized_residuals}
    R_i = \frac{y_i - y(\nu_i)}{\alpha_i},
\end{equation}
have a standard deviation closest to one. In Eq.~(\ref{eq:normalized_residuals}), $y_i$ and $y(\nu_i)$ denote the measured and fitted intensity at frequency $\nu_i$, respectively, and $\alpha_i$ is the corresponding uncertainty. Two noise sources contribute to the uncertainty $\alpha_i$: a constant background noise $\sigma_\mathrm{BG}$ described by a Gaussian distribution, and the shot noise $\sigma_i$ of the ion detection, which follows a Poisson distribution. The line shapes were first fitted in an unweighted manner to obtain an estimate $\tilde{y}(\nu_i)$ of $y(\nu_i)$, including a linear background term $y_\mathrm{BG}$. The uncertainty $\sigma_i$ of the He$_2$$^+$ signal strength was then set to $\sigma_i = \sqrt{\tilde{y}(\nu_i) - y_\mathrm{BG}}$. Finally, a second weighted fit was performed with the absolute weights set to $\alpha_i^{-2}$ with $\alpha_i = \sqrt{\sigma_i^2 + \sigma_\mathrm{BG}^2}$. Adding the Poissonian and Gaussian uncertainties in quadrature is an approximation that works best in the limit of large count numbers and was used to simplify the analysis.

To determine the transition frequencies, a pair of Voigt profiles was fitted to each fine-structure component by adjusting a common Lorentz width $\varGamma_{\mathrm{L},j}$, two line centers $\nu_j\pm\Delta_\mathrm{D}$, and two separate amplitudes $a_{j,1}$ and $a_{j,2}$. A single Doppler splitting $2\Delta_\mathrm{D}$ was fitted for all pairs of any given spectrum. This procedure was adequate whenever all lines were resolved. When this was not the case, e.g., for the two unresolved central components of the $c(4,2) \leftarrow a(0,1)$, $c(4,6) \leftarrow a(0,5)$, and $c(4,6) \leftarrow a(0,7)$ transitions, a common Lorentz width $\varGamma_{\mathrm{L}}$ was assumed for all lines, and the relative amplitudes of the unresolved components was constrained to follow Eq.~(\ref{eq:rel_intensities}). In the cases of the $c(4,2) \leftarrow a(0,3)$ and $c(4,8) \leftarrow a(0,9)$ transitions, one of the blue-shifted Doppler components of one transition overlapped with the red-shifted Doppler component of another transition. In these cases, an additional measurement was carried out after blocking the retroreflected cw-laser beam, resulting in spectra consisting solely of the red-shifted Doppler components. Their line shapes were then fitted by allowing individual Lorentz widths, amplitudes and peak positions to be adjusted. The final fit was then performed for the spectra consisting of both sets of Doppler components but using the Lorentz widths and fine-structure splittings from the previous fit and adjusting individual amplitudes for each line, a common Doppler shift $\Delta_\mathrm{D}$ and the center position of the lowest fine-structure component. The resulting wavenumbers of all observed transitions are listed in Table~\ref{tab:transitions}. The fits of the $2_3\leftarrow1_2$, $2_2\leftarrow1_1$, $2_1\leftarrow1_2$, $6_7\leftarrow5_6$, and $6_6\leftarrow5_5$ components turned out to be unreliable, and these transitions were not considered when deriving molecular constants.

\begin{figure}[b]
    \includegraphics[width=0.8\linewidth]{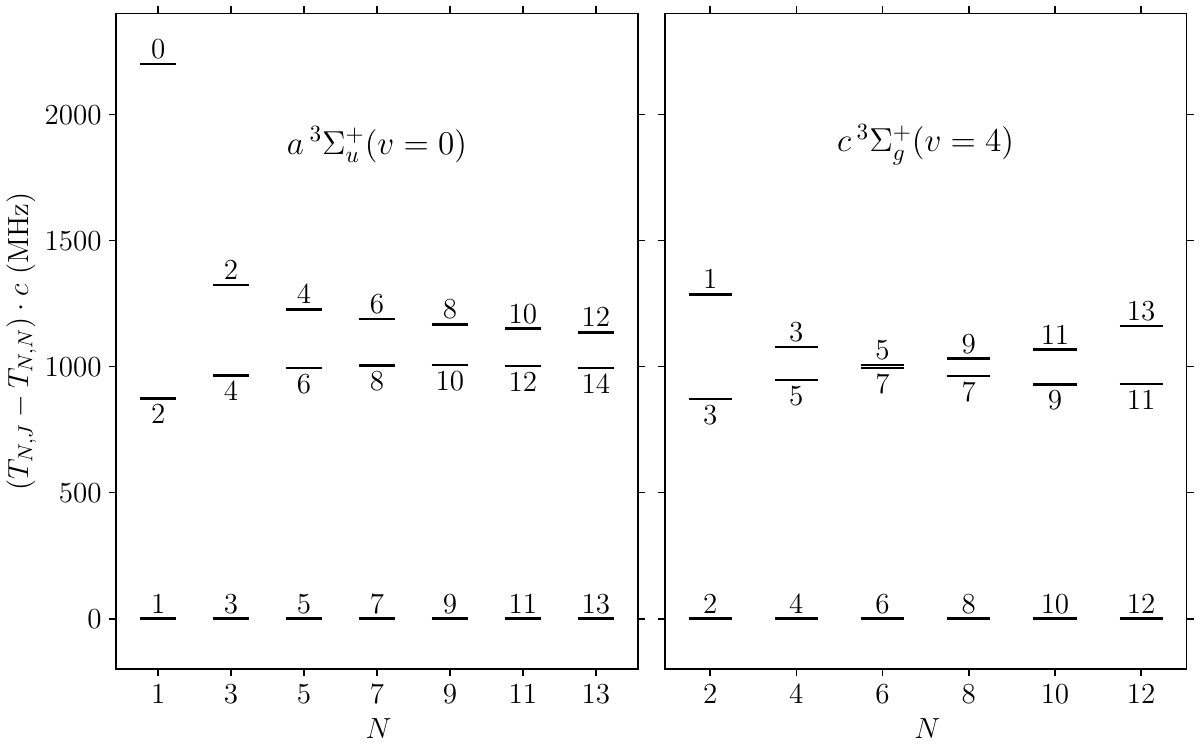}
    \caption{\label{fig:ac_splitting}\small Fine-structure intervals of the $a\;{}^3\Sigma_u^+(v=0)$ and $c\;{}^3\Sigma_g^+(v=4)$ states as a function of the rotational-angular-momentum quantum number $N$ relative to the $J=N$ fine-structure component. The relative term values of each fine-structure component are indicated as black bars with the total-angular-momentum quantum number $J$ above or below.}
\end{figure}

Because the data in Table~\ref{tab:transitions} represent a redundant set of $\mathcal{N}$ transition frequencies ${\boldsymbol{\nu}^T = [\nu_{1}, \nu_{2}, \ldots, \nu_{\mathcal{N}}]}$ between the $\mathcal{M}$ fine-structure components of the $a\;{}^3\Sigma_u^+(v=0)$ and $c\;{}^3\Sigma_g^+(v=4)$ states with term values $\boldsymbol{t}^T = [t_{1}, t_{2}, \ldots, t_{\mathcal{M}}]$, the term values $t_i$ relative to the position of the $a\;{}^3\Sigma_u^+({v^{\prime\prime}=0},{N^{\prime\prime}=1},{J^{\prime\prime}=1})$ level were fitted to the measured transition frequencies $\nu_j$ in a weighted linear least-squares fit using \cite{albritton76a} 
\begin{equation}\label{eq:linear_least_squares}
    \boldsymbol{\nu} = \boldsymbol{Xt} + \boldsymbol{\varepsilon}.
\end{equation}
In Eq.~(\ref{eq:linear_least_squares}), $\boldsymbol{\varepsilon}$ is an error and the elements of the matrix $X$ are
\begin{equation}
        X_{nm} = 
    \begin{cases}
      +1 & \text{if } N^{\prime}_{J^{\prime}}(\nu_n) = N^{\prime}_{J^{\prime}}(t_m),\\
      -1 & \text{if } N^{\prime\prime}_{J^{\prime\prime}}(\nu_n) = N^{\prime\prime}_{J^{\prime\prime}}(t_m),\\
       0 & \text{otherwise,}
    \end{cases}
\end{equation}
where $N_{J}(\nu_n)$ and $N_{J}(t_m)$ denote the $N$ and $J$ quantum numbers associated with the transition frequency $\nu_n$ and the term value $t_m$, respectively. The transitions were weighted using $(\sigma_n^\mathrm{stat} + \sigma^\mathrm{sys})^{-2}$, where $\sigma_n^\mathrm{stat}$ represents the statistical uncertainty of the transition frequency $\nu_n$ obtained from the line-shape fits and $\sigma^\mathrm{sys}$ was set to 200\,kHz for all transitions, except $c(4,0) \leftarrow a(0,1)$. To improve the quality of the fit, the fine-structure intervals of the $a\;{}^3\Sigma_u^+(v=0)$ state from Refs.~\citenum{hazell95a,semeria18a} and the transition frequencies between the $a\;{}^3\Sigma_u^+(v^{\prime\prime}=0,N^{\prime\prime}=7-13)$ and $c\;{}^3\Sigma_g^+(v^{\prime}=4,N^{\prime}=8-12)$ levels from Ref.~\citenum{kristensen90a} were included in the fit in addition to the transition frequencies measured in the present work. The resulting term values are listed in Table~\ref{tab:termvalues}. The fine-structure intervals of the $c\;{}^3\Sigma_g^+(v=4,N)$ levels evolve differently with the rotational-angular-momentum quantum number $N$ than those of the $a\;{}^3\Sigma_u^+(v=0)$ state and the fine-structure intervals of the different rotational levels of the $a\;{}^3\Sigma_u^+(v=0)$ and $c\;{}^3\Sigma_g^+(v=4)$ states are compared in Fig.~\ref{fig:ac_splitting}. For the $a\;{}^3\Sigma_u^+(v=0)$ state, the separation between the two upper components of the fine-structure triplet decreases as $N$ increases, while the energetic ordering of the levels is preserved. The energetic ordering of the fine-structure levels of the $c\;{}^3\Sigma_g^+(v=4,N=2)$ level is the same as that of the $a\;{}^3\Sigma_u^+(v=0)$ state. However, with increasing $N$, the separation between the upper components decreases, leading to nearly degenerate levels for $N=6$ and an inversion of the order for $N\geq8$. This behavior is the result of the different signs of the spin-rotation interaction in the $a\;{}^3\Sigma_u^+(v=0)$ and $c\;{}^3\Sigma_g^+(v=4)$ states (see further discussion in Section~\ref{sec:molecular_constants}).

\begin{table}[ht!]
    \renewcommand{\arraystretch}{0.75}
    \caption{\label{tab:termvalues}\small Term values $T_{N,J}$ relative to the ${a\;{}^3\Sigma_u^+(v=0,N=1,J=1)}$ state obtained from a weighted linear least-squares fit of the transition frequencies as well as transition frequencies reported in Refs.~\citenum{kristensen90a,semeria18a,hazell95a}. Numbers in parentheses denote 1$\sigma$ in units of the last digit.}
    \begin{minipage}[t]{0.48\linewidth}
    $a\;{}^3\Sigma_u^+(v=0)$\\
    \begin{ruledtabular}
    \begin{tabular}{r r D{.}{.}{4.14}}
        \multicolumn{1}{c}{$N$} & \multicolumn{1}{c}{$J$} & \multicolumn{1}{c}{$T_{N,J}$ (cm$^{-1}$)} \\
        \colrule
         1 &  1 &   0.0 \\
           &  2 &   0.029\,142\,481(15) \\
           &  0 &   0.073\,383\,149(12) \\
        \\
         3 &  3 &  75.813\,028(19) \\
           &  4 &  75.845\,217(19) \\
           &  2 &  75.857\,189(19) \\
        \\
         5 &  5 & 211.994\,106(22) \\
           &  6 & 212.027\,281(22) \\
           &  4 & 212.035\,036(22) \\
        \\
         7 &  7 & 408.061\,210(28) \\
           &  8 & 408.094\,730(28) \\
           &  6 & 408.100\,862(28) \\
        \\
         9 &  9 & 663.322\,383(33) \\
           & 10 & 663.355\,947(33) \\
           &  8 & 663.361\,292(33) \\
        \\
        11 & 11 & 976.882\,76(57) \\
           & 12 & 976.916\,19(57) \\
           & 10 & 976.921\,12(57) \\
        \\
        13 & 13 & 1347.6393(10) \\
           & 14 & 1347.6725(10) \\
           & 12 & 1347.6772(10) \\
    \end{tabular}
    \end{ruledtabular}
    \end{minipage}
    \hfill
    \begin{minipage}[t]{0.48\linewidth}
    $c\;{}^3\Sigma_g^+(v=4)$\\
    \begin{ruledtabular}
    \begin{tabular}{r r D{.}{.}{6.12}}
        \multicolumn{1}{c}{$N$} & \multicolumn{1}{c}{$J$} & \multicolumn{1}{c}{$T_{N,J}$ (cm$^{-1}$)} \\
        \colrule
         0 &  1 & 16\,069.447\,3626(29) \\
        \\
         2 &  2 & 16\,101.457\,807(15) \\
           &  3 & 16\,101.486\,897(24) \\
           &  1 & 16\,101.500\,691(12) \\
        \\
         4 &  4 & 16\,175.942\,131(21) \\
           &  5 & 16\,175.973\,690(22) \\
           &  3 & 16\,175.978\,085(21) \\
        \\
         6 &  6 & 16\,292.248\,981(30) \\
           &  7 & 16\,292.282\,144(41) \\
           &  5 & 16\,292.282\,545(25) \\
        \\
         8 &  8 & 16\,449.359\,857(33) \\
           &  7 & 16\,449.391\,973(32) \\
           &  9 & 16\,449.394\,259(34) \\
        \\
        10 & 10 & 16\,645.817\,355(66) \\
           &  9 & 16\,645.848\,370(50) \\
           & 11 & 16\,645.852\,937(67) \\
        \\
        12 & 12 & 16\,879.630\,63(95) \\
           & 11 & 16\,879.6617(10) \\
           & 13 & 16\,879.6693(11) \\
    \end{tabular}
    \end{ruledtabular}
    \end{minipage}
\end{table}

The term values of the fine-structure components of the $c\;{}^3\Sigma_g^+(v=4)$ state listed in Table~\ref{tab:termvalues} were used to extract pure experimental values for the fine-structure intervals of all $c\;{}^3\Sigma_g^+(v=4,N)$ rotational levels with $2\leq N \leq10$ for comparison with intervals determined very recently to high precision by R{\'a}csai \textit{et al.} \cite{racsai25a} in calculations including nonadiabatic, relativistic and leading-order QED corrections. The comparison, presented in Table~\ref{tab:splittings}, demonstrates a remarkable agreement. Most deviations are within the experimental uncertainties $\sigma_{ij}$, which were determined from the uncertainties of the individual term values $i$ and $j$ using $\sigma_{ij}^2 = \sigma_i^2 + \sigma_j^2 - 2r_{ij}\sigma_i\sigma_j$, where $r_{ij}$ is the correlation coefficient between the two term values. The inclusion of the leading-order QED corrections to the spin-dependent couplings in the calculations reported in Ref.~\citenum{racsai25a} was necessary to achieve this high level of agreement (see Table II of Ref.~\citenum{racsai25a}), which, in turn, implies that the experimental results are sufficiently precise to reveal such effects. 

\begin{table}[t!]
    \renewcommand{\arraystretch}{0.75}
    \caption{\label{tab:splittings}\small Comparison of the fine-structure intervals of the $N=2-10$ rotational levels of the $c\;{}^3\Sigma_g^+(v=4)$ state obtained from the term values listed in Table~\ref{tab:termvalues} (obs.) with the calculated intervals reported in Ref.~\citenum{racsai25a} (calc.). $\Delta=\mathrm{obs.}-\mathrm{calc.}$}
    \begin{ruledtabular}
    \begin{tabular}{ r r D{.}{.}{1.10} D{.}{.}{1.6} D{.}{.}{7.0}}
        & & \multicolumn{3}{c}{$T_{N,J} - T_{N,N}$ (cm$^{-1}$)} \\
        \cline{3-5} 
        \multicolumn{1}{c}{$N$} & \multicolumn{1}{c}{$J$} & \multicolumn{1}{c}{obs.} & \multicolumn{1}{c}{calc.} & \multicolumn{1}{c}{$\Delta\times10^6$}\\
        \colrule
         2 & 1 & 0.042\,884(18) & 0.042\,891 &  -7(18) \\
           & 3 & 0.029\,090(21) & 0.029\,123 & -33(21) \\
        \\
         4 & 3 & 0.035\,954(12) & 0.035\,942 &  12(12) \\
           & 5 & 0.031\,559(13) & 0.031\,597 & -38(13) \\
        \\
         6 & 5 & 0.033\,565(21) & 0.033\,546 &  19(21) \\
           & 7 & 0.033\,163(36) & 0.033\,185 & -22(36) \\
        \\
         8 & 7 & 0.032\,115(20) & 0.032\,072 &  43(20) \\
           & 9 & 0.034\,402(23) & 0.034\,461 & -59(23) \\
        \\
        10 & 9 & 0.031\,015(68) & 0.030\,948 &  67(68) \\
          & 11 & 0.035\,582(81) & 0.035\,600 & -18(81) \\
    \end{tabular}
    \end{ruledtabular}
\end{table}

\subsection{Decay Dynamics}\label{sec:decay_dynamics}
All rotational levels of the $c\;{}^3\Sigma_g^+(v=4)$ state are located above the $\mathrm{He(1\;^1S_0) + He(2\;^3S_1})$ dissociation limit. As a result, these levels are quasibound and do not only decay by fluorescence to the $a\;{}^3\Sigma_u^+$ state, but can also dissociate by quantum-mechanical tunneling through the $c$-state potential hump (see Fig.~\ref{Fig_potentials}). Whereas the tunneling lifetimes are expected to rapidly decrease with increasing $N$ values, they should not strongly depend on the fine structure. Hence, the centroid positions $T_{v,N}=\sum_Jw_JT_{v,N,J}$ with the statistical weights $w_J = (2J+1)/\sum_J(2J+1)$ of the experimental term values given in Table~\ref{tab:termvalues} were determined with respect to the $\mathrm{He(1s)^2\;{}^1S_0 + He(1s)(2s)\;{}^3S_1}$ dissociation limit using
\begin{align}
    E_{v,N}^\mathrm{kin}/(hc) &= T_{v,N} - D_0^{a(0,1)}, \nonumber\\
      &= T_{v,N} - E_{\mathrm{I}}^{a(0,1)}/(hc) - D_0^{X^+(0,1)} + E_{\mathrm{I}}^{\mathrm{He}(2\,{}^3S_1)}/(hc),
\end{align}
where $E_{v,N}^\mathrm{kin}$ represents the central position of the quasibound levels above the dissociation limit, $E_{\mathrm{I}}^{a(0,1)}/(hc) = 34\,301.207\,002(23)_\mathrm{stat}(37)_\mathrm{sys}\,\mathrm{cm}^{-1}$ the ionization energy of the $a\;{}^3\Sigma_u^+(v=0,N=1)$ state from Ref.~\citenum{semeria20a}, $D_0^{X^+(0,1)} = 19\,101.29(10)\,\mathrm{cm}^{-1}$ the dissociation energy of ${X^+\;{}^2\Sigma_u^+(v=0,N=1)}$ from Ref.~\citenum{holdener25a}, and $E_{\mathrm{I}}^{\mathrm{He}(2\,{}^3S_1)}/h = 1\,152\,842\,742.7082(55)_\mathrm{stat}(25)_\mathrm{sys}\,\mathrm{MHz}$ the ionization energy of metastable helium from Ref.~\citenum{clausen25a}. The results are listed in the second column of Table~\ref{tab:resonances}. The experimental resonance positions relative to the $c\;{}^3\Sigma_g^+(v=4,N=0)$ level (third column of Table~\ref{tab:resonances}) are compared to the \textit{ab-initio} calculations of Ref.~\citenum{racsai25a} in the fourth column of Table~\ref{tab:resonances} and show excellent agreement.

Additionally, the resonance positions and widths of the $c\;{}^3\Sigma_g^+(v=4,N)$ states were calculated by determining the phase-shift of the nuclear radial wavefunction as a function of energy, as described in Ref.~\citenum{holdener25a}. We used a potential-energy function originally employed in Ref. \citenum{jordan86a}. This function is divided into three regions, ensuring continuity of $V(R)$ and $\mathrm{d}V(R)/\mathrm{d}R$ over all region boundaries: a short-range region (A) with a well representing the bond, a middle-range region (B) with a hump representing the barrier, and a long-range region (C) where electrostatic forces become dominant (see Ref.~\citenum{jordan86a} for details). This function was refined in Ref.~\citenum{lorents89a} based on the rovibrational-level positions known at that time and assuming a $c$-state dissociation energy of $D_0^{c(0,0)}=4023.82\,\mathrm{cm}^{-1}$ estimated from a value of $19\,073\pm50\,\mathrm{cm}^{-1}$ for the dissociation energy of the $X^+\;{}^2\Sigma_u^+$ state of ${}^4$He$_2$. A similar approach was followed here but considering the more accurate value of the dissociation energy $D_0^{X^+(0,1)}$ of the $X^+(0,1)$ ground state of ${}^4$He$_2$$^+$ mentioned above, which implies a dissociation energy of $D_0^{c(0,0)} = 4073.50(10)\,\mathrm{cm}^{-1}$ for the $c\;{}^3\Sigma_g^+(0,0)$ state of ${}^4$He$_2$. Because the analysis of Ref.~\citenum{lorents89a} reproduced the $c$-state rovibrational term values accurately, we only needed to shift the potential in regions A and B by a constant offset $\Delta V$, leaving the long-range part C unchanged. By solving the Schr{\"o}dinger equation for the $c(0,0)$ level with the log-derivative method \cite{johnson77a}, a value of $\Delta V/(hc)=-50.97\,\mathrm{cm}^{-1}$ was obtained. The absolute resonance positions above the dissociation limit obtained with the shifted potential are compared to the experimental values in the sixth column of Table~\ref{tab:resonances}, and show good agreement with the experimental values. However, the relative energies are not as accurate as those of the \textit{ab-initio} calculations of Ref.~\citenum{racsai25a}.

\begin{figure}[b]
    \includegraphics[width=0.6\linewidth]{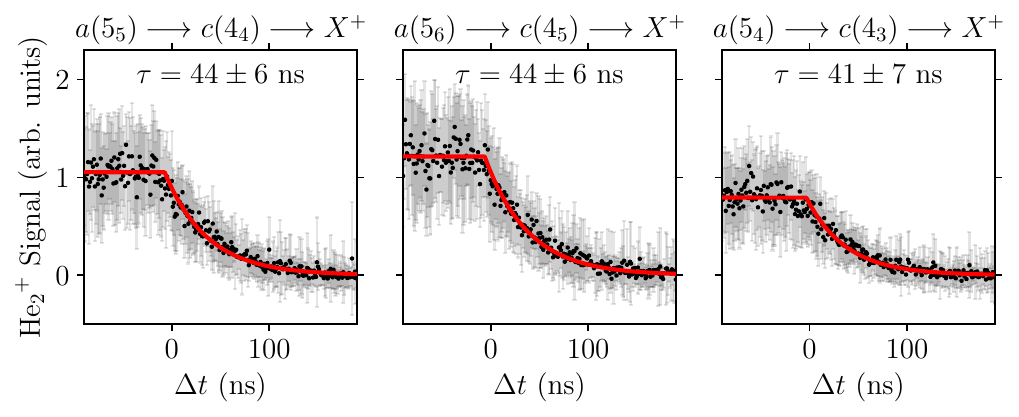}
    \caption{\label{fig:P5_decay}\small Determination of the lifetimes of the $c\;{}^3\Sigma_g^+(v=4,N=4,J=3,4,5)$ states by monitoring the decay of the He$_2$$^+$ signal generated by ($1+1^\prime$) R2PI as a function of the delay between the excitation and ionization laser pulses. The gray dots with error bars represent the average and corresponding standard deviation of 20 individual measurement cycles. The red lines correspond to the fitted exponential decay.}
\end{figure}

\begin{table*}[t!]
    \renewcommand{\arraystretch}{0.75}
    \caption{\label{tab:resonances}\small Observed resonance positions of rotational levels of the $c\;{}^3\Sigma_g^+(v=4)$ state of ${}^4$He$_2$ relative to the $\mathrm{He(1\;^1S_0) + He(2\;^3S_1})$ dissociation limit ($E_\mathrm{kin}$) and to the $c\;{}^3\Sigma_g^+(v=4,N=0)$ level ($E_\mathrm{rot}$). The experimental values obtained in this work are compared to \textit{ab-initio} calculations from Ref.~\citenum{racsai25a} ($\Delta E_\mathrm{rot} = E_\mathrm{rot}^\mathrm{obs.} - E_\mathrm{rot}^\mathrm{calc.}$) and to calculations based on a model potential adapted from Ref.~\citenum{lorents89a} ($\Delta E_\mathrm{kin} = E_\mathrm{kin}^\mathrm{obs.} - E_\mathrm{kin}^\mathrm{calc.}$). For the model potential, the tunneling widths $\varGamma$ (FWHM) have also been calculated.}
    \begin{ruledtabular}
    \begin{tabular}{r D{.}{.}{4.3} D{.}{.}{3.10} D{.}{.}{3.2} D{.}{.}{2.2} D{.}{.}{2.2} D{.}{.}{1.8}}
        & \multicolumn{3}{c}{observed} & \multicolumn{3}{c}{calculated} \\
        \cline{2-4} \cline{5-7}
        & \multicolumn{2}{c}{This Work} & \multicolumn{1}{c}{Ref.~\citenum{lorents89a}} & \multicolumn{1}{c}{Ref.~\citenum{racsai25a}} & \multicolumn{2}{c}{Modified from Ref.~\citenum{lorents89a}} \\
        \cline{2-3} \cline{4-4} \cline{5-5} \cline{6-7}
        $N$  & \multicolumn{1}{c}{$\frac{E_\mathrm{kin}}{hc}$ / cm$^{-1}$} & \multicolumn{1}{c}{$\frac{E_\mathrm{rot}}{hc}$ / cm$^{-1}$} & \multicolumn{1}{c}{$\frac{E_\mathrm{rot}}{hc}$ / cm$^{-1}$} & \multicolumn{1}{c}{$\frac{\Delta E_\mathrm{rot}}{hc}$ / cm$^{-1}$} & \multicolumn{1}{c}{$\frac{\Delta E_\mathrm{kin}}{hc}$ / cm$^{-1}$} & \multicolumn{1}{c}{$\varGamma$ / MHz}\\
        \colrule
         0 & 1121.62\footnote{Values in this column have an uncertainty of 0.10\,cm$^{-1}$ from the dissociation energy $D_0^{c(0,0)}$.} &    0.00         &   0.00 &  0.00 & -1.17 & 5.07\times10^{-2} \\
         2 & 1153.65 &   32.032\,596(25) &  31.97 & -0.01 & -1.15 & 7.51\times10^{-2} \\
         4 & 1228.14 &  106.516\,947(32) & 106.35 & -0.01 & -1.11 & 1.78\times10^{-1} \\
         6 & 1344.44 &  222.823\,840(45) & 222.58 & -0.03 & -1.04 & 6.51\times10^{-1} \\
         8 & 1501.56 &  379.934\,757(47) & 379.71 & -0.04 & -0.93 & 3.38 \\
        10 & 1698.01 &  576.392\,336(74) & 576.31 & -0.07 & -0.80 & 2.29\times10^{1} \\
        12 & 1931.83 &  810.2067(13)  & 810.26 & -0.07 & -0.59 & 1.86\times10^{2} \\
    \end{tabular}
    \end{ruledtabular}
\end{table*}

The spin-rotational levels of the $c\;{}^3\Sigma_g^+(v=4)$ state can decay both by tunneling predissociation and fluorescence to the $a\;{}^3\Sigma_u^+$ state. Their lifetimes $\tau$ are thus given by ${{\tau}^{-1} = \tau_\mathrm{rad}^{-1} + \tau_\mathrm{tun}^{-1}}$, where $\tau_\mathrm{rad}$ and $\tau_\mathrm{tun}$ are the radiative and tunneling lifetimes, respectively. To obtain experimental values of $\tau$ and the corresponding Lorentzian widths $\varGamma$, the decay of the $c\;{}^3\Sigma_g^+(v=4,N=4,J)$ fine-structure levels was measured in a pump-probe experiment by monitoring the decrease of the He$_2$$^+$ signal after turning off the cw radiation and delaying the ionization pulse, as explained in Section~\ref{sec:experimental}. The results are presented in Fig.~\ref{fig:P5_decay}, which reveals exponentially decaying signals after the cw radiation is turned off. A weighted nonlinear least-squares fit of an exponential decay starting at a fitted time $t_0$ yielded a lifetime of 43(6)\,ns for the $c\;{}^3\Sigma_g^+(v=4,N=4)$ level, which corresponds to a natural linewidth of 3.7(6)\,MHz. This width is much larger than the calculated tunneling predissociation widths for levels with $N<8$ (see Table~\ref{tab:resonances}). Consequently, the decay of rotational levels with $N<8$ is entirely dominated by the radiative decay of the $c\;{}^3\Sigma_g^+$ state. The predissociation becomes the main decay channel for levels with $N\geq10$. The radiative lifetime of the $c\;{}^3\Sigma_g^+(v=4,N=4)$ level was calculated to be 72\,ns, following the procedure outlined in Ref.~\citenum{verdegay25a} and using the potential-energy functions and electric-dipole transition moment of Ref.~\citenum{yarkony89a}. This calculation disregarded $c$-state wavefunction amplitudes beyond the barrier maximum as well as contributions from emission to the $a$-state continuum. The theoretical lifetime is significantly longer than the experimentally determined one (43(6)\,ns). This discrepancy might be the result of the approximation mentioned above. 

The lines of the $c(4,10) \leftarrow a(0,9)$ transition are significantly broader than all other transitions involving $c\;{}^3\Sigma_g^+(v=4,N<10)$ levels. The Lorentzian widths obtained from the lineshape fits are 25.6(12)\,MHz, 25.0(18)\,MHz and 27.5(16)\,MHz for the $10_{9}\leftarrow9_{8}$, $10_{10}\leftarrow9_{9}$ and $10_{11}\leftarrow9_{10}$ transitions, respectively. These values are about twice as large as the 13.5\,MHz Gaussian residual Doppler linewidth and are in good agreement with the calculated resonance tunneling width of 22.9\,MHz (see Table~\ref{tab:resonances}), if the measured 3.7(6)\,MHz width from the radiative decay is added.

\subsection{Molecular Constants}\label{sec:molecular_constants}
The fine-structure levels of ${}^3\Sigma$ states can be described with an effective Hamiltonian, consisting of rotational, spin-rotation and spin-spin terms \cite{brown79a}
\begin{equation}
\begin{aligned}\label{eq:effectiveHamiltonian}
    H_\mathrm{eff} = &B_v\boldsymbol{N}^2 - D_v\boldsymbol{N}^4 + H_v\boldsymbol{N}^6 + \ldots\\
    &+\left(\gamma_v + \gamma_{D_v}\boldsymbol{N}^2 + \gamma_{H_v}\boldsymbol{N}^4 + \ldots\right)\boldsymbol{N}\cdot\boldsymbol{S}\\
    &+\left[\lambda_v + \lambda_{D_v}\boldsymbol{N}^2 + \lambda_{H_v}\boldsymbol{N}^4 + \ldots,{S_z}^2-\frac{1}{3}\boldsymbol{S}^2\right]_+,
\end{aligned}
\end{equation}  
where $[\cdot,\cdot]_+$ denotes an anticommutator. In He$_2$, the states are well described in Hund's angular-momentum-coupling case (b) with the angular-momentum-coupling scheme $\mathbf{J} = \mathbf{N} + \mathbf{S} = \mathbf{R} + \mathbf{L} + \mathbf{S}$, where $\mathbf{J}$ is the total angular momentum, $\mathbf{S}$ the total electron-spin angular momentum, and $\mathbf{N}=\mathbf{R}+\mathbf{L}$ the sum of the rotational ($\mathbf{R}$) and the electronic orbital angular momentum ($\mathbf{L}$). In this coupling scheme, the rotational and spin-rotational part of Eq.~(\ref{eq:effectiveHamiltonian}), i.e., the first two lines, are diagonal in $N$ and $J$. The spin-spin part [third line of Eq.~(\ref{eq:effectiveHamiltonian})] has both diagonal and off-diagonal matrix elements. The sums $\alpha_J^N$ of diagonal terms for a fine-structure level with given $N$ and $J$ values are given by \cite{brown03a}
\begin{equation}\label{eq:diagoanl_elements}
    \alpha_J^N = \begin{cases}
        B_{v,N}N(N+1)-\frac{2}{3}\frac{N}{2N+3}\lambda_{v,N} + N\gamma_{v,N}, & J=N+1,\\
        B_{v,N}N(N+1)+\frac{2}{3}\lambda_{v,N} - \gamma_{v,N}, & J=N,\\
        B_{v,N}N(N+1)-\frac{2}{3}\frac{N+1}{2N-1}\lambda_{v,N} - (N+1)\gamma_{v,N}, & J=N-1,\\
    \end{cases}
\end{equation}
where the rotational, spin-spin and spin-rotational constants undergo centrifugal-distortion effects according to \cite{brown79a}
\begin{equation}
\begin{aligned}
    B_{v,N} &= B_{v} - D_{v}N(N+1) + H_{v}N^2(N+1)^2 + \ldots, \\
    \lambda_{v,N} &= \lambda_{v} + \lambda_{D_v}N(N+1) + \lambda_{H_v}N^2(N+1)^2 + \ldots, \\
    \gamma_{v,N} &= \gamma_{v} + \gamma_{D_v}N(N+1) + \gamma_{H_v}N^2(N+1)^2 + \ldots .
\end{aligned}
\end{equation}
The spin-spin interaction couples fine-structure components according to the selection rules $\Delta J = 0$ and $\Delta N = 2$. We denote the corresponding off-diagonal matrix elements as $\beta_J$ and they are given by \cite{brown03a}
\begin{equation}\label{eq:off_diagonal_elements}
    \beta_J = \frac{\sqrt{J(J+1)}}{2J+1}\big(\lambda_{v,J-1}+\lambda_{v,J+1}\big).
\end{equation}
The Hamiltonian in Eq.~(\ref{eq:effectiveHamiltonian}) is block-diagonal and can be analytically diagonalized to obtain the energy eigenvalues
\begin{equation}\label{eq:FS_energies}
    E_J^N = \begin{cases}
        \alpha_J^N, & N=J\text{ or }J=0,\\
        \frac{1}{2}\left[\Sigma_J \pm \left\{\beta_J^2+\Delta_J^2\right\}^{1/2}\right], & N=J\pm1,
    \end{cases}
\end{equation}
where $\Sigma_J = \alpha_J^{J+1} + \alpha_{J}^{J-1}$ and $\Delta_J = \alpha_J^{J+1} - \alpha_{J}^{J-1}$.

\begin{table*}[ht!]
\caption{\label{tab:constants_a}\small Comparison of the molecular constants of the $a\;{}^3\Sigma_u^+(v=0)$ state of ${}^4$He$_2$ determined in this work with the results of previous investigations \cite{semeria18a,focsa98a,hazell95a,kristensen90a,rogers88a,ginter65a}. All values are given in cm$^{-1}$.}
\begin{ruledtabular}
\tiny
\begin{tabular}{l D{.}{.}{2.13} D{.}{.}{2.13} D{.}{.}{2.13} D{.}{.}{2.12} D{.}{.}{2.9} D{.}{.}{2.12} D{.}{.}{1.8}}
 & \multicolumn{1}{c}{This Work} & \multicolumn{1}{c}{Ref.~\citenum{semeria18a}} & \multicolumn{1}{c}{Ref.~\citenum{focsa98a}} & \multicolumn{1}{c}{Ref.~\citenum{hazell95a}} & \multicolumn{1}{c}{Ref.~\citenum{kristensen90a}} & \multicolumn{1}{c}{Ref.~\citenum{rogers88a}} & \multicolumn{1}{c}{Ref.~\citenum{ginter65a}} \\
\colrule
$B_0$                        &  7.589\,1610(10)                                         & \multicolumn{1}{c}{$-$}                  &  7.589\,141(27)   & \multicolumn{1}{c}{$-$}               & \multicolumn{1}{c}{$-$}             &  7.589\,199(26)  & 7.5863(10) \\
$D_0\times10^4$              &  5.618\,44(25)                                           & \multicolumn{1}{c}{$-$}                  &  5.615\,29(135)   & \multicolumn{1}{c}{$-$}               & \multicolumn{1}{c}{$-$}             &  5.6228(28)      & 5.50(5) \\
$H_0\times10^8$              &  3.395(17)                                               & \multicolumn{1}{c}{$-$}                  &  3.217(25)        & \multicolumn{1}{c}{$-$}               & \multicolumn{1}{c}{$-$}             &  3.47(10)        & 2.5(4) \\
$L_0\times10^{12}$           & -3.480\footnote{Fixed to value of Ref.~\citenum{focsa98a}.} & \multicolumn{1}{c}{$-$}                  & -3.480(130)       & \multicolumn{1}{c}{$-$}               & \multicolumn{1}{c}{$-$}             & -6.1(11)         & \multicolumn{1}{c}{$-$}\\
\\
$\lambda_0\times10^2$        & -3.666\,437\,36(48)                                      & -3.666\,437\,00(40) & -3.666\,4342(128) & -3.666\,4328(67) & -3.666\,436(6) & -3.666\,4333(84) & \multicolumn{1}{c}{$-$}\\
$\lambda_{D_0}\times10^6$    &  6.587\,06(44)                                           &  6.5872(17)         &  6.5887(37)       &  6.5867(21)      &  6.5896(30)    & 6.5852(46)       & \multicolumn{1}{c}{$-$}\\
$\lambda_{H_0}\times10^{10}$ & -0.1558(18)                                              & -0.1561(80)         & -0.1595(94)       & -0.1544(57)      & -0.187(23)     & \multicolumn{1}{c}{$-$}               & \multicolumn{1}{c}{$-$}\\
$\lambda_{L_0}\times10^{14}$ &  4.47(17)                                                &  4.50(77)           &  4.65(62)         &  4.37(43)        & \multicolumn{1}{c}{$-$}             & \multicolumn{1}{c}{$-$}               & \multicolumn{1}{c}{$-$}\\
\\
$\gamma_0\times10^5$         & -8.079\,31(39)                                           & -8.079\,72(20)      & -8.0805(22)       & -8.0804(12)      & -8.0793(20)    & -8.0751(38)      & \multicolumn{1}{c}{$-$}\\
$\gamma_{D_0}\times10^8$     &  2.2843(20)                                              &  2.2869(33)         &  2.2828(70)       &  2.2839(70)      &  2.268(22)     & 2.02(15)         & \multicolumn{1}{c}{$-$}\\
$\gamma_{H_0}\times10^{12}$  & -1.979(21)                                               & -2.011(47)          & -1.943(62)        & -1.955(80)       & \multicolumn{1}{c}{$-$}             & \multicolumn{1}{c}{$-$}               & \multicolumn{1}{c}{$-$}\\
\end{tabular}
\end{ruledtabular}
\end{table*}

\begin{table*}[ht!]
\caption{\label{tab:constants_c}\small Comparison of the molecular constants of the $c\;{}^3\Sigma_g^+(v=4)$ state of ${}^4$He$_2$ determined in this work with the results of previous investigations \cite{kristensen90a,li10a,ginter65a,lorents89a}. All values are given in cm$^{-1}$.}
\begin{ruledtabular}
\begin{tabular}{l D{.}{.}{6.12} D{.}{.}{3.7} D{.}{.}{6.12} D{.}{.}{6.9} D{.}{.}{3.8}}
 & \multicolumn{1}{c}{This Work} & \multicolumn{1}{c}{Ref.~\citenum{kristensen90a}} & \multicolumn{1}{c}{Ref.~\citenum{li10a}} & \multicolumn{1}{c}{Ref.~\citenum{ginter65a}} & \multicolumn{1}{c}{Ref.~\citenum{lorents89a}}\\
\colrule
$T_4$                     & 16\,084.599\,1100(31)    & \multicolumn{1}{c}{$-$}         & 16\,084.6134(24)    & 16\,084.610(10)\footnote{Estimated from Table IV of Ref.~\citenum{ginter65a}.} & \multicolumn{1}{c}{$-$}\\
\\
$B_4$                     &       5.344\,276\,25(94) & \multicolumn{1}{c}{$-$}         &       5.344\,36(21) &       5.3400(10)                                                            &   5.333(3) \\
$D_4\times10^4$           &       9.164\,63(37)      & \multicolumn{1}{c}{$-$}         &       9.190(47)     &       8.57(4)                                                               &   7.647(206)\\
$H_4\times10^8$           &     -23.904(58)          & \multicolumn{1}{c}{$-$}         &     -19.7(36)       &     -54.9(20)                                                               & -81.5(40)\\
$L_4\times10^{14}$        &      -4.838(28)          & \multicolumn{1}{c}{$-$}         &      -6.87(88)      & \multicolumn{1}{c}{$-$}                                                                          & \multicolumn{1}{c}{$-$}\\
\\
$\lambda_4\times10^2$     &      -3.266\,64(82)      & -3.176(63) & \multicolumn{1}{c}{$-$}                  & \multicolumn{1}{c}{$-$}                                                                          & \multicolumn{1}{c}{$-$}\\
$\lambda_{D_4}\times10^6$ &      -2.66(23)           & -16.0(53)  & \multicolumn{1}{c}{$-$}                  & \multicolumn{1}{c}{$-$}                                                                          & \multicolumn{1}{c}{$-$}\\
\\
$\gamma_4\times10^5$      &      36.31(14)           & 35.82(50)  & \multicolumn{1}{c}{$-$}                  & \multicolumn{1}{c}{$-$}                                                                          & \multicolumn{1}{c}{$-$}\\
$\gamma_{D_4}\times10^8$  &       2.7(23)            & 23.0(43)   & \multicolumn{1}{c}{$-$}                  & \multicolumn{1}{c}{$-$}                                                                          & \multicolumn{1}{c}{$-$}\\
\end{tabular}
\end{ruledtabular}
\end{table*}

The $c\;{}^3\Sigma_g^+(v=4)-a\;{}^3\Sigma_u^+(v=0)$ band origin and the rotational, spin-spin and spin-rotation constants of the $a\,{}^3\Sigma_u^+({v=0})$ and $c\,{}^3\Sigma_g^+(v=4)$ states including their respective centrifugal-distortion corrections were determined in a global weighted nonlinear least-squares fit based on Eq.~(\ref{eq:FS_energies}). The fit included transition frequencies from Refs.~\citenum{kristensen90a,semeria18a,hazell95a} in addition to the data summarized in Table~\ref{tab:transitions}. The results of the fit are listed in Table~\ref{tab:constants_a} for the $a\;{}^3\Sigma_u^+(v=0)$ state and Table~\ref{tab:constants_c} for the $c\;{}^3\Sigma_g^+(v=4)$ state, where they are compared to earlier results \cite{semeria18a,focsa98a,hazell95a,kristensen90a,lorents89a,ginter65a,kristensen90a,li10a}. Overall, the agreement with previously published results is very good, but the present results represent a significant improvement in the precision of most molecular constants. The band origin of the $c\;{}^3\Sigma_g^+(v=4) - a\;{}^3\Sigma_u^+(v=0)$ transition is found to be 300\,MHz lower than earlier values. This deviation is likely to lie in the calibration of the absolute frequency. Whereas an optical frequency comb referenced to a Rb atomic-clock standard was used in the present work, a wavemeter was used in Ref.~\citenum{li10a}, in combination with I$_2$ reference lines, resulting in an absolute accuracy of about $0.005\,\mathrm{cm}^{-1}\approx150\,\mathrm{MHz}$.

The values of the zeroth-order spin-spin constant $\lambda_v$ are similar in both electronic states, but the spin-rotational constant $\gamma_v$ of the $c\;{}^3\Sigma_g^+(v=4)$ state has a magnitude approximately four times larger than that of the $a\;{}^3\Sigma_u^+(v=0)$ state and the opposite sign. This difference is probably caused by nonadiabatic and spin-orbit interactions of the $c\;{}^3\Sigma_g^+$ with the close-lying $b\;{}^3\Pi_g$ state. The nonadiabatic interactions couple the electronic angular momentum with the rotational angular momentum, and the spin-orbit interaction couples the electronic angular momentum with the spin angular momentum, which in turn also affects the coupling of the electron spin with the molecular rotation. 

In a first-order approximation, the off-diagonal elements of the effective Hamiltonian [Eq.~(\ref{eq:off_diagonal_elements})] can be neglected. The energetic ordering of the fine-structure levels can then be directly inferred from Eq.~(\ref{eq:diagoanl_elements}). The splitting in the fine-structure levels with $J=N\pm1$ is then given by
\begin{equation}
    \Delta\alpha^N_{N\pm1} = \frac{4N+2}{4N(N+1)-3}\lambda_{v,N} + (2N + 1)\gamma_{v,N}.
\end{equation}
If $\lambda_{v,N}$ and $\gamma_{v,N}$ have the same sign, the relative energetic ordering is the same for all $N$. However, if the two fine-structure constants have opposite signs, the order changes at 
\begin{equation}
    N(N+1) = \frac{1}{2}\left|\frac{\lambda_{v,N}}{\gamma_{v,N}}\right| + \frac{3}{4}.
\end{equation}
Using the zeroth-order spin-spin and spin-rotation constants of the $c\;{}^3\Sigma_g^+(v=4)$ state given in Table~\ref{tab:constants_c}, we obtain $N=6.3$, which implies that the ordering should change between $N=6$ and $N=8$, as observed experimentally (see Table~\ref{tab:termvalues} and Fig.~\ref{fig:ac_splitting}).

\section{Conclusions}
We have presented a precision measurement of the ${c\;{}^3\Sigma_g^+(v=4) - a\;{}^3\Sigma_u^+(v=0)}$ band of ${}^4$He$_2$ using single-mode continuous-wave laser radiation. Calibration of the frequency with an optical frequency comb referenced to a Rb-GPS frequency standard enabled the determination of the frequencies of 40 transitions with a relative accuracy in the range $2.5\times10^{-10}<\Delta\nu/\nu<2.8\times10^{-9}$ (see Table~\ref{tab:transitions}). 

By selecting narrow transverse-velocity classes of the ${}^4$He$_2$ molecules in the molecular beam using a resonant two-photon ionization scheme, linewidths as narrow as 15\,MHz (full width at half maximum) were achieved, allowing for the first time the full resolution of the fine-structure of the $N=0,2,4,6,8$ and $10$ rotational levels of the $c\;{}^3\Sigma_g^+(v=4)$ state, and the detection of a sharp decrease of tunneling lifetimes at $N=10$. The experimental fine-structure intervals of the $c\;{}^3\Sigma_g^+(v=4)$ state are found to be in excellent agreement with recent highly precise \textit{ab-initio} calculations \cite{racsai25a}. 

By combining the results presented in this work with previous measurements, the molecular constants of the $a\;{}^3\Sigma_u^+(v=0)$ and $c\;{}^3\Sigma_g^+(v=4)$ could be improved significantly in a global nonlinear least-squares fit. 

Measurements of the radiative and tunneling decay dynamics of the quasi-bound $c\;{}^3\Sigma_g^+(v=4,N=4,10)$ levels enabled the determination of radiative and tunneling-predissociation linewidths of 3.7(3)\,MHz and 22.9\,MHz, respectively. The tunneling-predissociation linewidths could be quantitatively accounted for in calculations using a model potential-energy function for the $c$ state.

\begin{acknowledgments}
We thank Josef A. Agner and Hansjürg Schmutz for technical support and Edit Mátyus for useful discussions and for providing results prior to publication. This work is supported financially by the Swiss National Science Foundation under project 200021-236716.
\end{acknowledgments}

%

\end{document}